\documentclass[12pt]{article}
\usepackage{amsmath}
\usepackage{cite}
\usepackage{epsfig}
\usepackage{epstopdf}
\usepackage{amssymb}

\def\[{\left [}
\def\]{\right ]}
\def\({\left (}
\def\){\right )}

\def\r{\rho}

\def\r2{\sqrt{2}}

\newcommand{\bbibitem}[1]{\bibitem{#1}\marginpar{#1}}

\def\Label#1{\label{#1}%
  \smash{\hbox to0pt{\raise1ex\hbox{\tiny[#1]}\hss}}}
\def\noLabels{\let\Label=\label}
\def\nobbibitem{\let\bbibitem=\bibitem}

\newcommand{\bea}{\begin{eqnarray}}
\newcommand{\eea}{\end{eqnarray}}

\newcommand{\beq} {\begin{equation}}
\newcommand{\eeq} {\end{equation}}
\newcommand{\beqa} {\begin{eqnarray}}
\newcommand{\eeqa} {\end{eqnarray}}

\newcommand{\dr}{\delta \tilde{R}}
\newcommand{\da}{\delta A_t}
\newcommand{\db}{\delta A_\theta}

\newcommand{\beqn}{\begin{eqnarray}}
\newcommand{\eeqn}{\end{eqnarray}}

\begin{document}

\begin{flushright}
HIP-2009-32/TH
\end{flushright}

\vskip 2cm \centerline{\Large {\bf Inhomogeneous Structures in
    Holographic Superfluids:}}
 \vskip 2mm
\centerline{\Large {\bf II. Vortices}}
 \vskip 1cm
\renewcommand{\thefootnote}{\fnsymbol{footnote}}
\centerline
{{\bf Ville Ker\"anen,$^{1,2}$
\footnote{ville.keranen@helsinki.fi}
Esko Keski-Vakkuri,$^{1}$\footnote{esko.keski-vakkuri@helsinki.fi}
Sean Nowling,$^{1,3}$\footnote{sean.nowling@helsinki.fi}
K. P. Yogendran$^{1,4}$ \footnote{yogendran.kalpat@helsinki.fi}
}}
\vskip .5cm
\centerline{\it
${}^{1}$Helsinki Institute of Physics }
\centerline{\it P.O.Box 64, FIN-00014 University of
Helsinki, Finland}
\centerline{\it
${}^{2}$Department of Physics}
\centerline{\it P.O.Box 64, FIN-00014 University of
Helsinki, Finland}
\centerline{\it
${}^{3}$Department of Mathematics and Statistics}
\centerline{\it P.O.Box 68, FIN-00014 University of
Helsinki, Finland}
\centerline{\it
${}^{4}$IISER Mohali}
\centerline{\it MGSIPAP Complex, Sector 26
   Chandigarh 160019 , India}

\setcounter{footnote}{0}
\renewcommand{\thefootnote}{\arabic{footnote}}

\begin{abstract}
We study vortex solutions in a holographic model of Herzog, Hartnoll,
and Horowitz, with a vanishing external magnetic field on the
boundary, as is appropriate for vortices in a superfluid. We study the
relevant length scales related to the vortices and how the charge
density inside the vortex core behaves as a function of temperature or
chemical potential. We extract a critical superfluid velocity from the
vortex solutions, study how it behaves as a function of the
temperature, and compare it to earlier studies and to the Landau
criterion. We also comment on the possibility of a
Berezinskii-Kosterlitz-Thouless vortex confinement-deconfinement
transition.

\end{abstract}

\newpage

\section{Introduction}
One of the focal points of activity in applying holographic methods to
condensed matter systems has been studies of holographic
superconductors. An initial spark was the observation in
\cite{Gubser:2008px}, where it was noted that an Abelian Higgs model
coupled to gravity in AdS space exhibits spontaneous symmetry breaking
of the local $U(1)$ symmetry (in the bulk) as the charge of the black
hole was increased. In \cite{Hartnoll:2008vx}, this phase transition
was interpreted in the dual field theory as a superconducting phase
transition, since in the model a charged operator gets a vacuum
expectation value (VEV).

In this paper we take a more conservative stance and use the
superfluid interpretation \cite{Herzog:2008he,Basu:2008st}.  We are
continuing our investigation of inhomogeneous extended configurations
in the model \cite{Hartnoll:2008vx}, which we begun in
\cite{Keranen:2009vi,Keranen:2009ss} reporting on holographical dark
solitons. In this paper we show that the model \cite{Hartnoll:2008vx}
allows vortex solutions even in the {\em absence} of a magnetic field
on the boundary.  In this case, the superfluid interpretation is most
natural: in a superconductor a vortex would act as a source for a
dynamical magnetic field and require turning on a flux line, whereas
we are explicitly setting the boundary $\vec{B}=0$ and yet find a
consistent solution.
Droplet and vortex solutions in the presence of boundary magnetic
fields have previously been discussed in
\cite{Albash:2009iq,Montull:2009fe,Maeda:2009vf}.\footnote{For other studies of external magnetic fields on holographic superconductors see for example \cite{Albash:2008eh,Hartnoll:2008kx,Albash:2009ix,Zeng:2009dv}.} An early work on
holographic vortices is a construction of a vortex line in pure AdS
space \cite{Dehghani:2001ft} and in the AdS-Schwarzschild background
\cite{Mann}.

We construct vortex solutions in \cite{Hartnoll:2008vx}, by solving
the equations of motion of the Abelian Higgs model in the probe
approximation, i.e. neglecting the backreaction of the scalar field and
the gauge fields on the black hole geometry.  We study basic
properties of the vortex solutions, such as the free energy and
associated length scales. We also study the amount of density
depletion in the core of the vortex, and find agreement with earlier
studies \cite{Keranen:2009vi,Keranen:2009ss}. In \cite{Keranen:2009ss}
we noted that the density depletion in the core of a dark soliton
depends on the dimension of the condensing operator in a way
reminiscent to a loosely bound fermion pair for dimension $\Delta=2$
and a more tightly bound fermion pair for $\Delta=1$, by comparing to
an earlier study in non-relativistic superfluids
\cite{antezzaetal}. For vortices a similar study of the depletion
fraction in a non-relativistic setting is in \cite{Randeria}.

In this work we extend the analogy between the type of the superfluid
and the dimension of the condensing operator by studying critical
superfluid velocities above which the superfluid flow starts to
dissipate and superfluidity is ruined. The critical velocity is
defined from the core size of the vortices. Our results for critical
flows partially overlap with others obtained from constant superfluid
flow in \cite{Herzog:2008he,Basu:2008st}.

We also compare the critical velocity of superfluid flow to velocity
of second sound (the second sound velocity was calculated in \cite{Herzog:2008he,Amado:2009ts} using different techniques), in the
spirit of Landau criterion, and find that for both scaling dimensions
the superfluid velocity is below the sound velocity. The proposal that
superfluidity in the $\Delta=2$ case might be due to loosely bound
fermions is partly supported by the fact that the critical velocity is
not set by the superfluid's second sound.

The outline of the paper goes as follows. In section
\ref{sec:Goldstone} we introduce the model and use cylindrical
symmetry to derive the form of the vortex solution, and the
corresponding equations of motion. In sections \ref{sec:model} and
\ref{sec:method} we describe our numerical methods of solving the
partial differential equations.  In sections \ref{sec:scales} and
\ref{sec:depletion} we study the characteristic length scales of the
vortex and the density depletion. Section \ref{sec:velocity} discusses
the critical velocity as determined from the vortices.  In section
\ref{sec:bkt} we show that the free energy of one vortex is $\log$
divergent as is the case in superfluids (a related calculation has
been done in \cite{Montull:2009fe}), and we study how the coefficient
of the log term behaves as a function of temperature and comment on
the possibility of a Berezinskii-Kosterlitz-Thouless (BKT) phase
transition \cite{KT,B} in the system.  Finally we end with a
discussion and outline future directions.

\section{Review of superfluids and vortices}\label{sec:Goldstone}
We will begin by reviewing some basic properties of superfluids in 2+1
dimensions and how vortices arise in them and what role they play. A
superfluid is characterized by spontaneous breaking of a global $U(1)$
symmetry. This already poses a theoretical puzzle in $2+1$ dimensions
since such spontaneous symmetry breaking at finite temperature should
not be possible due to the Mermin-Wagner theorem. In the holographic
model, this problem is perhaps being circumvented by having a large
number $N$ of degrees of freedom in the field theory
\footnote{See additional discussion in \cite{Horowitz:2008bn,Herzog:2009xv,Hartnoll:2009sz}.
The question of identifying the $N$ degrees of freedom can be
addressed after embedding holographic model into string/M theory see
\cite{Gubser:2009qm,Gauntlett:2009dn,Gubser:2009gp,Gauntlett:2009bh,Denef:2009tp}
for work in this direction.}. In this section, we will assume that
spontaneous symmetry breaking is stabilized in the large-$N$ limit and
we will look at the consequences of that for the low energy physics of
the superfluid.

The low lying excitations in a superfluid are the Goldstone bosons
from spontaneous breaking of the $U(1)$ symmetry. At large length
scales, as compared to the inverse mass of the lightest massive
quasiparticle, the system is well described by an effective action for
the Goldstone field $\phi(x)$. At zero chemical potential where the
effective action is manifestly Lorentz invariant, the only
non-irrelevant operator one can write down is
$(\partial_{\mu}\phi)^2$. Generalizing this to a finite chemical
potential, $\mu$, can be done by introducing a formal gauge field
$A_{\mu}$ with one non-zero component, $A_0=\mu$
\cite{Son:2002zn}. Since the effective action respects the formal
gauge invariance $\phi\rightarrow e^{i\Lambda}\phi$,
$A_{\mu}\rightarrow A_{\mu}+\partial_{\mu}\Lambda$ it has the form
\beq
S_{eff}=-\int d^3x\frac{1}{2}\kappa(D_{\mu}\phi)^2=\int
d^3x\frac{1}{2}\kappa\Big((\partial_0\phi-\mu)^2
-(\partial_i\phi)^2\Big),\label{eq:eff}
\eeq
The action (\ref{eq:eff}) has a conserved momentum density
$T_{0\mu}=\kappa\partial_{0}\phi\partial_{\mu}\phi$ and a current
density $j_{i}=\kappa\partial_{i}\phi$. Both of the four vectors are
parallel to $\partial_{\mu}\phi$, which motivates one to define the
superfluid velocity vector $v_s^{\mu}=\gamma\partial^{\mu}\phi$. The
coefficients $\kappa$ and $\gamma$ seem at the moment to be arbitrary,
but for a relativistic superfluid they can be connected to
thermodynamic quantities as \cite{Valle}
\beq
\kappa=\frac{\rho_s}{\mu},\quad\gamma=\frac{1}{\mu}\ ,\label{eq:const}
\eeq where $\rho_s$ is the superfluid density.
These relations will be important in later sections, so that we will
derive them in a simple example in the next section.

We can define a vortex as a state for which the expectation value of
the superfluid velocity $\langle
v_s^{i}\rangle=\langle\partial^{i}\phi\rangle/\mu$ is cylindrically
symmetric and has a non-zero circulation
\beq
c=\mu\oint dx_{i}\langle v_s^{i}\rangle,\label{eq:circu}
\eeq
where the integral is taken over circle of constant radius. Using the
averaged (time independent) Heisenberg equation
\beq
0=\langle\partial_i\partial^{i}\phi\rangle=\mu\partial_i \langle v_s^{i} \rangle,\label{eq:Heis}
\eeq
we see that the radial part of the superfluid velocity is a
constant. We will focus on the case where this constant
vanishes. Using (\ref{eq:circu}) we get the superfluid velocity of the
vortex as
\beq
\langle v_s^{i}\rangle=c\epsilon^{ij}\frac{x^{j}}{\rho^2},\label{eq:supvel}
\eeq
where $\rho$ is the radial coordinate on two dimensions. For a normal
fluid the circulation $c$ can take any values, but in a superfluid
$\phi$ is the phase of a complex field and thus, the circulation is
quantized in integer multiplets of $2\pi$. Quantized circulation is
one of the hallmarks of superfluid vortices.

Even though it is easy to see the existence of vortices simply by
topological arguments and/or by use of the simple Goldstone effective
action, the interior of the vortex is model dependent and carries
interesting information about the microscopic structure of the
superfluid \cite{antezzaetal}. Since the superfluid velocity diverges
inside the vortex core, the vortex solution must interpolate all the
way from the superfluid to the normal phase, with $\rho_s=0$, as a
function of $\rho$. Thus, in order to describe a vortex, one is forced
to appeal to a microscopic description that is able to describe the
normal phase.

The energy of a single vortex can be estimated from the effective
action (\ref{eq:eff}) as
\beq
E_{vor}\approx\int d^2x\frac{1}{2}\mu\rho_s(\langle v_s\rangle)^2
\approx\frac{\pi\rho_s n^2}{\mu}\textrm{log}\Big(\frac{R_c}{\xi}\Big),
\eeq
where $R_c$ is an IR cutoff and $\xi$ a UV cutoff. In a microscopic
description the UV divergence is absent since $\rho_s$ vanishes at the
vortex core, so we are left with an IR divergence in the vortex
energy. This does not still mean that vortices would not exist in
superfluids. Vortices will be formed in a superfluid when the system
is put under rotation with angular frequency $\omega$, which is larger
than a critical frequency $\omega_c$ (see eg. \cite{Dalfovo}). If the
size of the system is taken to infinity, the critical angular velocity
tends to zero and vortices will immediately form when the system is
rotating. Another scenario where vortices are important is during the
BKT transition.  In that situation, the entropy of a vortex becomes
comparable to $E_{vor}/T$ and it becomes thermodynamically preferred
to nucleate vortices in the system.

The divergence of the vortex energy is in accordance with Derrick's
theorem \cite{Derrick}, which states that the energy of a soliton
solution in a theory with only scalar fields (with canonical kinetic
terms) is divergent if the space dimension of the system $D$ is larger
than 1. In the holographic setting, there are also gauge fields
involved, so that Derrick's theorem as such does not apply, and thus
it is not \textit{a priori} clear, whether the vortex energy is finite
or not, in the holographic model. But as we will see, the vortex
energy is indeed logarithmically divergent (for a related calculation
see \cite{Montull:2009fe}). This is necessary for the superfluid phase
to exist in non-zero temperature, since if it would take finite energy
to create a vortex, it would be entropically favorable at any
temperature to nucleate a soup of vortices, which would drive the
order parameter to zero.
\subsection{An example model with vortices: The Gross-Pitaevskii equation}\label{sec:GP}
To get some insight into the microscopic structure of superfluid
vortices, before attacking the holographic problem, we study vortices
in the relativistic Gross-Pitaevskii (GP) equation (we will drop the
label "relativistic" in what follows). The GP equation can be thought
of as arising from a saddle point evaluation of a path integral for
the bosonic field $\Psi$, with the grand canonical action
\beq
S_{GP}=\int d^3x\Big(|\partial_0\Psi-i\mu\Psi|^2-|\partial_i\Psi|^2-V|\Psi|^2-\frac{1}{2}g|\Psi|^4\Big).\label{eq:GPaction}
\eeq
The GP equation is relevant for weakly interacting, dilute
Bose-Einstein (BEC) condensates and tightly bound fermionic
superfluids at low temperatures.

It is interesting to see how one can end up on an effective action of
the form (\ref{eq:eff}) from the GP theory. This gives a simple
example where one can derive the identifications (\ref{eq:const})
easily.  By integrating out the fluctuations of the modulus in
(\ref{eq:GPaction}) we end up with an effective action for the
Goldstone field , of the form
\beq
S_{eff}=\int d^3x |\Psi_0|^2\Big(|\partial_0\phi-\mu|^2-|\partial_i\phi|^2\Big),
\eeq
where $\Psi_0$ is the VEV of $\Psi$. This action is clearly of the
form (\ref{eq:eff}). In the GP equation the superfluid density is
simply the particle number density derived from the action
(\ref{eq:GPaction}),
$\rho_s=\frac{-i}{2}(\Psi^{*}\partial_t\Psi-\Psi\partial_t\Psi^{*}+2i\mu|\Psi|^2)\approx\mu|\Psi_0|^2$,
where in the second equality we have expanded around the ground state
and ignored fluctuations. Thus the effective action becomes
\beq
S_{eff}=\int d^3x\frac{\rho_s}{\mu}\Big(|\partial_0\phi-\mu|^2-|\partial_i\phi|^2\Big),
\eeq
which is the same thing as (\ref{eq:eff}) with
$\kappa=\rho_s/\mu$. The particle number current is now
$j_i\approx|\Psi_0|^2\partial_i\chi=\rho_s\partial_i\chi/\mu\equiv\rho_s
v_s$, which tells us that $v_s=\partial_i\chi/\mu$. Thus we see how
the identification (\ref{eq:const}) arises from relativistic GP
theory.

\begin{figure}
\begin{center}
\includegraphics[scale=0.7]{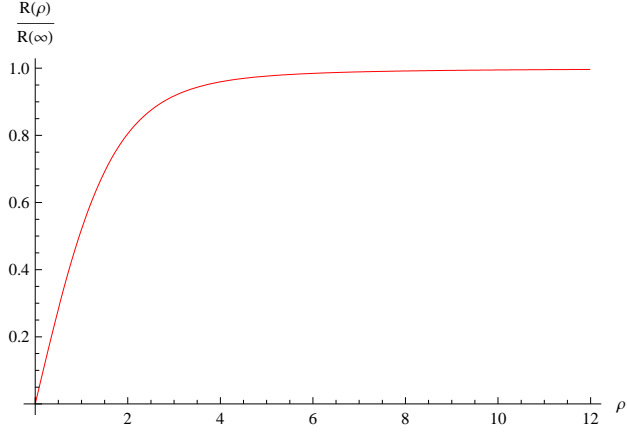}
\caption{\label{fig:GPfit} Spatial profile of a vortex solution to the
  GP equation.}
\end{center}
\end{figure}

Let us proceed to discuss vortices. For a time independent field
configuration the GP equation is
\beq
-\nabla^2\Psi+(V-\mu^2)\Psi+g|\Psi|^2\Psi=0.\label{eq:GPeq}
\eeq
Note that the non-relativistic version of the time independent GP
equation can be obtained from (\ref{eq:GPeq}) simply by replacing
$\mu^2$ by $\mu$. An \textit{ansatz} for a vortex solution can be
taken in the form $\Psi=e^{in\theta}R(\rho)$, where $n$ is the winding
number (or circulation) of the vortex. Plugging the \textit{ansatz}
into the GP equation gives
\beq
-R''-\frac{1}{\rho}R'+\frac{n^2}{\rho^2}R+(V-\mu^2)R+gR^3=0.\label{eq:GP}
\eeq
Again the superfluid velocity has the same form as in
(\ref{eq:supvel}). Since (\ref{eq:GP}) does not have known analytic
solutions for non-zero winding $n$, we will solve it numerically. The
numerical solution is displayed in Fig \ref{fig:GPfit}.
From the numerics it seems that the field $R(r)$ behaves in a power
law fashion at large values of $\rho$. This suggests that we can try
to get an analytic solution for asymptotically in $\rho$. Plugging a
power series \textit{ansatz} to (\ref{eq:GP}) gives the leading
behavior
\beq
R(\rho)=\sqrt{\frac{\mu^2-V}{g}}\Big(1-\frac{n^2\xi_2^2}{2\rho^2}+...\Big),\label{eq:asymp}
\eeq
and the particle number density (which is equivalent to the superfluid density) is given by
\beq
\rho_s(\rho)=\mu R^2(\rho)=\mu\frac{\mu^2-V}{g}\Big(1-\frac{n^2\xi_2^2}{\rho^2}+...\Big),
\eeq
where $\xi_2=1/\sqrt{\mu^2-V}$. For the unit winding vortex we can
also define a length scale from the vortex's core, $R\approx
R(\infty)(\rho/\xi_1+...)$.  We can numerically calculate $\xi_1$,
which gives the ratio
\beq
\frac{\xi_1}{\xi_2}\approx 1.75.\label{eq:ratio}
\eeq
Note that there is really only one length scale in the problem.  This
means that at sufficiently low temperatures (that is, when the GP
equation is valid) the ratio (\ref{eq:ratio}) is a constant as the
chemical potential is varied.  We will revisit this in the context of
holographic superfluids in section \ref{sec:scales}.

\section{The model}\label{sec:model}
As argued in \cite{Hartnoll:2008vx}, a holographic dual of a
superfluid is provided by a scalar field coupled to an
Einstein-Maxwell system in asymptotically $AdS$ space.
\beq
S_{AdS}=\int d^4x\sqrt{-g}\Big[\frac{1}{\kappa_4^2}\left(\mathcal{R}
-\frac{6}{L^2}\right)-\frac{1}{q^2}\Big(|D_{\mu}\Psi|^2
+m^2|\Psi|^2+\frac{1}{4}F_{\mu\nu}^2\Big)\Big],
\eeq
As in \cite{Hartnoll:2008vx}, will choose the mass
$m^2=-2/L^2$. Throughout this work we will work in the probe limit,
($\frac{\kappa_4^2}{q^2}$ is small), so that the backreaction to
gravity can be ignored. Thus, we are working with the fixed bulk
metric
\beq
ds^2=\frac{L^2}{z^2}(-f(z)dt^2+f(z)^{-1}dz^2+d\rho^2+\rho^2d\theta^2),
\eeq
where $f(z)=1-\frac{z^3}{z_T ^3}$. The AdS-CFT dictionary tells us that
\beq
e^{-S_{AdS}(on shell)}=\langle e^{-\int d^3x
  (\rho(x)\mu(x)+\mathcal{O}_1(x)\mathcal{O}_2(x))}\rangle_{QFT},
\eeq
where the boundary quantum field theory operators are related to the
boundary values of the bulk AdS fields by the relations
\beq
\Psi(x,z)=z\mathcal{O}_1(x)+z^2\mathcal{O}_2(x)+...\quad
A_{t}(x,z)=\mu(x)+z\rho(x)+...
\eeq
Thus, in order to find the QFT operator expectation values we need to
solve the classical equations of motion in AdS space to obtain the on
shell fields. Ignoring gravitational backreaction, the equations of
motion become
\begin{align}
0 &= \frac{1}{\sqrt{-g}}\partial_\mu(\sqrt{-g} g^{\mu\nu}\partial_\nu
R)+m^2 R - R(\partial_\mu\chi-A_\mu)^2\label{eq:ELeq1}
\\
0 &= \frac{1}{\sqrt{-g}}\partial_\mu(\sqrt{-g} F^{\mu\nu}) -
R^2(A^\nu-\partial^\nu\chi)\label{eq:ELeq2}
\\
0&=\partial_{\mu}(\sqrt{-g}R^2g^{\mu\nu}(\partial_{\nu}\chi-A_{\nu})),\label{eq:ELeq3}
\end{align}
where we have defined the real valued fields $R$ and $\chi$ according
to the relation $\Psi=\frac{1}{\sqrt{2}}Re^{i\chi}$.

The equations of motion may be greatly simplified using the
observation that cylindrical symmetry implies that gauge invariant
quantities are independent of $\theta$.  A detailed discussion of
gauge fixing and the use of symmetry is presented in Appendix
\ref{sec:gauge}.  An important point discussed in Appendix
\ref{sec:appendix1} is that the field $\chi$ is a normalizable mode.
If $\chi$ was non-normalizable, its boundary value would be set by
hand, which in turn would imply that features such as the superfluid
velocity profile are not determined by the system dynamically.

Working in the $A_z=0$ gauge and defining the field
$\tilde{R}=R/z$, the equations which describe a vortex profile are
\begin{align}
0&= f\partial_z^2\tilde{R}+\partial_zf\partial_z\tilde{R}-z\tilde{R}+
\frac{1}{\rho}\partial_\rho(\rho\partial_\rho \tilde{R})\nonumber
\\
&-\tilde{R}(-\frac{1}{f}A_t^2+\frac{(A_\theta-n)^2}{\rho^2})\label{eq:eqr}
\\
0 &= f\partial^2_z A_t+\frac{1}{\rho}\partial_\rho(\rho\partial_\rho
A_t)-\tilde{R}^2A_t\label{eq:eqt}
\\
0&= \partial_z( f\partial_z
A_\theta)+\rho\partial_\rho(\frac{1}{\rho}\partial_\rho A_\theta)
- \tilde{R}^2(A_\theta-n),\label{eq:eqs}
\end{align}
where all the fields are functions of only $z$ and $\rho$.  As
discussed in appendix \ref{sec:gauge} the only non-zero gauge field
components are $A_t$ and $A_\theta$.
\footnote{ Since there is a magnetic flux $\int_{z=z_0} F\neq 0$ going
  through the vortex core in the bulk, one can ask what happens to it
  as one approaches the boundary. Since the magnetic flux is conserved
  $\int_{\Sigma} F = 0$ for any closed surface $\Sigma$, the magnetic
  flux has to go somewhere as one approaches the boundary. Since we
  have a vanishing boundary magnetic field the magnetic flux does not
  reach the boundary, but rather it turns to the $\rho$ direction in
  the form of the magnetic field $B_{\rho}\sim 1/\rho$. In the
  holographic dictionary $B_{\rho}=\partial_{z}A_{\theta}/\rho$ is not
  identified as a magnetic field, but rather with the superfluid
  current $j_{\theta}$. So, in short, the magnetic flux turns into
  superfluid flow on the boundary.}

\section{Finding the vortex solutions}\label{sec:method}
\subsection{The method}\label{sec:GS}
The equations of motion are a set of coupled nonlinear partial
differential equations, which do not seem to be solvable analytically.
In order to proceed, we therefore resort to numerical methods.
Because the vortex solutions are inhomogeneous, it is difficult to use
standard differential equation solvers, such as Mathematica's NDSolve.
Instead we will use a Gauss-Seidel relaxation scheme.
\footnote{In Appendix \ref{sec:asy} we discuss a complementary
  approach valid away from the vortex core.  In that regime, the
  differential equations reduce to ordinary differential equations and
  one may use "off the shelf" tools.}  As described in
\cite{Keranen:2009vi,Keranen:2009ss}, we first place the system on a
finite box of radius $L$, $(z,\rho)\in[0,1]\times[0,L]$ and discretize
along both axes. After the usual procedure of replacing derivatives
with finite difference derivatives, the differential equations are
turned into finite difference equations.

The resulting finite difference equations are solved by choosing an
initial seed configuration and iterating with the Gauss-Seidel
method. The error in solving the equations of motion falls off with a
power law as the lattice size is increased. With a sufficient number
of iterations and a fine enough lattice, the seed configuration
relaxes to a solution to the equations of motion. For further details
on this numerical method see \cite{Keranen:2009ss}.

\subsection{Boundary conditions}
To find a unique solution to the finite difference equations, we
specify boundary conditions at each edge of the region being
simulated. The boundary conditions at the AdS boundary determine the
external sources turned on in the dual field theory. The gauge field,
$A_t$, takes a constant value (independent of $\rho$ and $\theta$) on
the boundary corresponding to a constant chemical potential,
$A_t(z=0)=\mu$. The scalar field, $\tilde{R}=|\Psi|/(\sqrt{2}z)$, can
satisfy either Dirichlet $\tilde{R}(z=0)=0$ or Neumann
$\partial_z\tilde{R}(z=0)=0$ boundary conditions, corresponding to two
different boundary theories with different scaling dimensions of the
condensing scalar operator, either $\mathcal{O}_2$ or
$\mathcal{O}_1$. Finally the angular gauge field, $A_{\theta}$,
satisfies Dirichlet boundary conditions, $A_{\theta}(z=0)=0$. This is
because we want the boundary magnetic field to vanish, implying
$\partial_{\rho}A_{\theta}(z=0)=0$. Because the boundary value of
$A_{\theta}$ is simply a constant, we can choose such a gauge that
$A_{\theta}(z=0)=0$.

On the $\rho=L$ boundary of the region being simulated we impose
Neumann boundary conditions for all the fields, since we want the
fields to asymptote to the translationally invariant symmetry breaking
solution.  We have verified that the solutions and results in this
paper are unchanged when one increases $L$, hence we can safely
conclude that we have used a large enough box.

Since the core of the vortex should be in the normal phase we impose
Dirichlet boundary conditions on the scalar field
$\tilde{R}(\rho=0)=0$ in the core.  Because the scalar field vanishes
in the origin, there is no bulk electric charge at the vortex core.
Therefore, along the $\rho=0$ boundary we impose Neumann boundary
conditions for the gauge field $A_{t}$ such that center of the vortex
is not a source for the radial electric field. This is basically a
regularity condition. Finally we impose the Dirichlet boundary
condition $A_{\theta}(\rho=0)=0$ on the angular gauge field. This may
also be regarded as a regularity conditions along the vortex core. We
want the bulk magnetic field $B_z= \partial_{\rho}A_{\theta}/\rho$ to
be finite at the origin. This means that $A_{\theta}$ has the behavior
$A_{\theta}=A_{\theta}^{(0)}(z)+\rho^2A_{\theta}^{(1)}(z)$ near the
vortex core. To have finite superfluid current in the core we require
the $\partial_z A_{\theta}/\rho$ to be finite in the limit
$\rho\rightarrow 0$. This means the $\partial_z
A_{\theta}^{(0)}=0$. Furthermore, since we required
$A_{\theta}(z=0)=0$ fixes the constant part $A_{\theta}^{(0)}=0$ and,
thus $A_{\theta}(\rho=0)=0$.

\begin{figure}[h]
\begin{center}
\includegraphics[height=4.5cm]{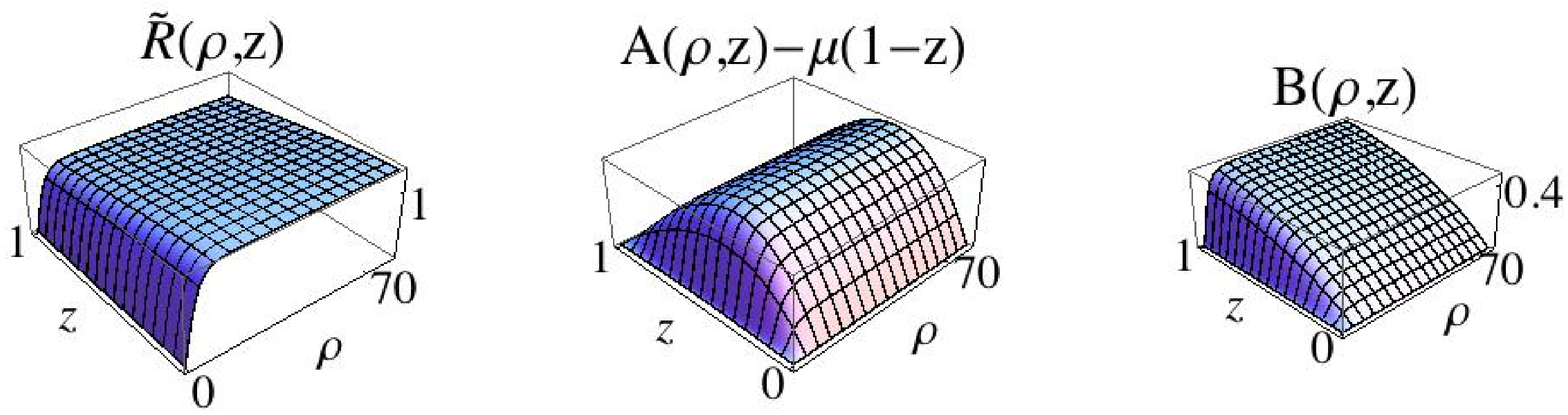}

\caption{\label{fig:O13d} Typical field profiles for the
  $<\mathcal{O}_1>$ condensate. (For visualization purposes we have
  subtracted a linear background from $A$.)}
\end{center}
\end{figure}

\begin{figure}[h]
\begin{center}
\includegraphics[height=4.5cm]{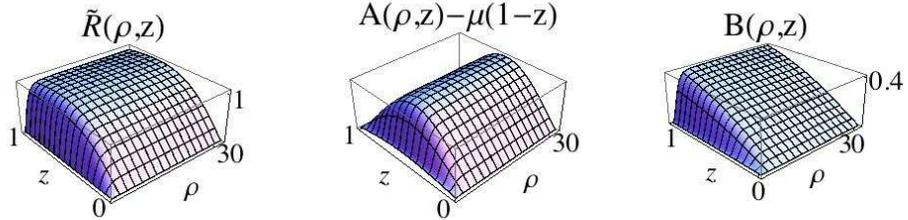}
\caption{\label{fig:O23d} Typical field profiles for the
  $<\mathcal{O}_2>$ condensate.  (For visualization purposes we have
  subtracted a linear background from $A$.)}

\end{center}
\end{figure}

The final boundary conditions to discuss are regularity conditions to
be imposed along the black hole horizon.  The equations of motion are
elliptic outside the horizon, but become parabolic along the horizon.
Also, we require that $A_t(z=1)=0$ so that the connection one form be
normalizable at the horizon. This leads to the following regularity
conditions at the horizon.
\begin{align}
0&=
(\partial_zf)\partial_z\tilde{R}-\tilde{R}+\frac{1}{\rho}\partial_\rho(\rho\partial_\rho
\tilde{R})-\tilde{R}\frac{(A_\theta-n)^2}{\rho^2}\label{eq:beqr} \\ 0
&= A_t\label{eq:beqa} \\ 0&= (\partial_z f)\partial_z
A_\theta+\rho\partial_\rho(\frac{1}{\rho}\partial_\rho A_\theta)-
\tilde{R}^2(A_\theta-n).\label{eq:beqs}
\end{align}

\subsection{Existence and stability}
Vortex solutions of (\ref{eq:eqr})-(\ref{eq:eqs}) are topologically
stable, with a conserved topological charge
\beq
n=\int_0^{2\pi}d\theta\partial_{\theta}\chi.
\eeq
For all regular solutions $n$ is an integer, and thus continuous time
evolution cannot change its value.

If one prepares the system in a configuration of non-zero winding, the
time evolution will keep the field in the same winding sector for all
times. Eventually the system will stabilize to the minimum energy
field configuration which solves the equations of motion with
vanishing time derivatives. They are simply the static equations
(\ref{eq:eqr}),(\ref{eq:eqt}), and (\ref{eq:eqs}). The minimum energy
field configuration is approximated by the solutions we find
numerically.

We are solving the equations using a relaxation method, which means
that the solution evolves in the iteration time $\tau$ by a
(generalized) diffusion equation. In order to establish that our
numerical solutions approach the minimum energy solution in the
desired winding sector, we need to show that the field $\tilde{R}$
does not generate a discontinuity to "unwind" the solution. The fact
that this does not happen follows from the diffusive nature of the
iteration algorithm. The diffusive nature guarantees that the energy
of a configuration is non-increasing. Thus, if we start from a seed
configuration with finite energy density above the ground state, we
will not generate discontinuities to the fields since they would
require too much energy.

\section{Characteristic scales}\label{sec:scales}

\begin{figure}[h]
\begin{center}
\includegraphics[scale=0.77]{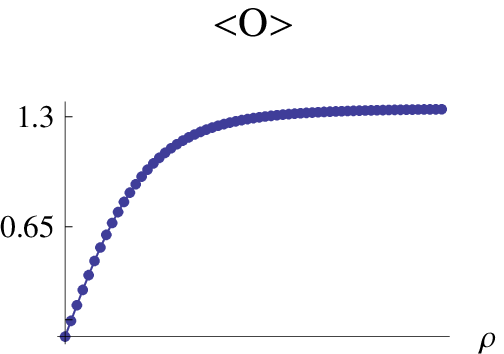}
\includegraphics[scale=0.77]{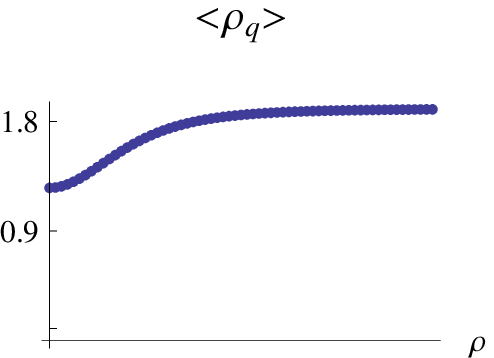}
\includegraphics[scale=0.77]{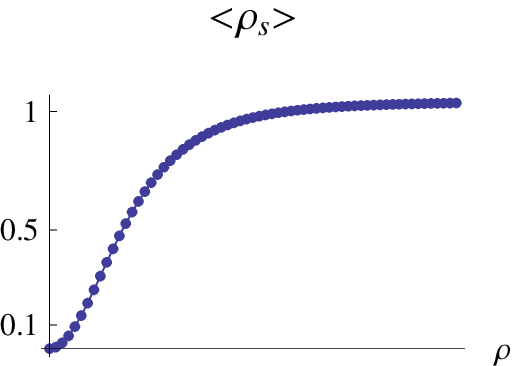}
\caption{\label{fig:fitsarray} Typical radial profiles for expectation
  values.}
\end{center}
\end{figure}

Typical condensate, charge density, and superfluid density radial
profiles as extracted from the bulk solutions, are shown in
Fig.(\ref{fig:fitsarray}).

We begin our analysis of the vortices described in section
(\ref{sec:GS}) by focusing on unit winding vortices.  The first
quantities we would like to identify in these vortices are the relevant
scales of variation.  It can be seen from our numerical solutions that
the fields have a power-law fall off as $\rho\rightarrow\infty$.  This
fact is confirmed in the tail regime as is discussed in appendix
\ref{sec:asy}.

Using the standard AdS/CFT dictionary, the behavior of the condensate
may be read off from the boundary behavior of the scalar field. In
this way we can define a coherence length by either fitting a
power-law function of the form
\beq
\langle\mathcal{O}_i(\rho)\rangle=e^{i\theta}|\langle\mathcal{O}_i(\infty)
\rangle|(1-\frac{\xi^2_2}{2\rho^2}+...)
,\label{eq:asycond}
\eeq
to the tail of the condensate field or from the slope of the
condensate in the vortex core.
\beq
\langle\mathcal{O}_i(\rho)\rangle=e^{i\theta}|\langle\mathcal{O}_i(\infty)\rangle|
\frac{\rho}{\xi_1}+...
\eeq
As discussed in Appendix \ref{sec:asy}, in the large $\rho$ regime the
differential equations may also be solved using shooting methods.  The
length scale in (\ref{eq:asycond}) may be determined with high
accuracy. We will conventionally define the length scales
as was done for solutions of the GP equation \ref{sec:GP}.

The boundary value of the derivative of the gauge field $A_{t}$ is
identified with the boundary theory charge density $\rho_q$ (or the
total particle number density in the superfluid language). It is seen
that it can be fitted well to a function of the form
\beq
\rho_q(\rho)=\rho_q(0)+(\rho_q(\rho)-\rho_q(0))(1-\frac{\xi_q^2}{\rho^2}+...),\label{eq:asyrho}
\eeq
with a characteristic charge density coherence length $\xi_q$.

The boundary value of the derivative of the angular gauge field
$A_{\theta}$ can be identified with a current $j_{\theta}$ in the
boundary theory. We will define the superfluid density $\rho_s$
through the relation
\beq
j^{i}= \rho_s v^{i}_s.
\eeq
Since $\chi(z=0)$ is the phase of the operator expectation value
\beq
\langle \mathcal{O}_i \rangle= \tilde{R}_ie^{i\chi(z=0)},
\eeq
we can identify the superfluid velocity as
\beq
v^{i}_s=\frac{1}{\mu}\partial^i\chi(z=0)=\frac{1}{\mu}\frac{n\varepsilon^{ij} x^{j}}{\rho^2},
\eeq
where the normalization factor follows from relativistic symmetry
\cite{Valle} as we mentioned earlier. Now the superfluid density for
the vortex can be identified as
\beq
\rho_s=\frac{\mu \partial_z A_{\theta}(z=0)}{n}.
\eeq
\begin{figure}
\begin{center}
\includegraphics[scale=0.85]{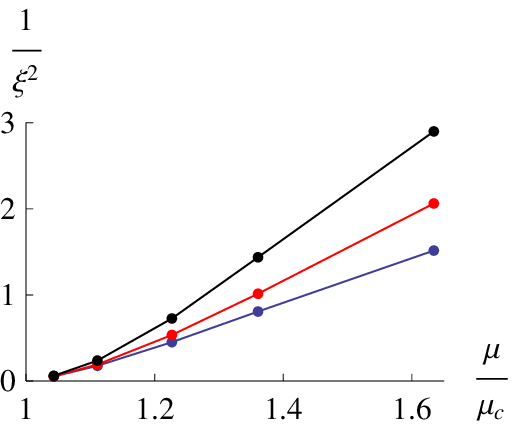}
\includegraphics[scale=0.85]{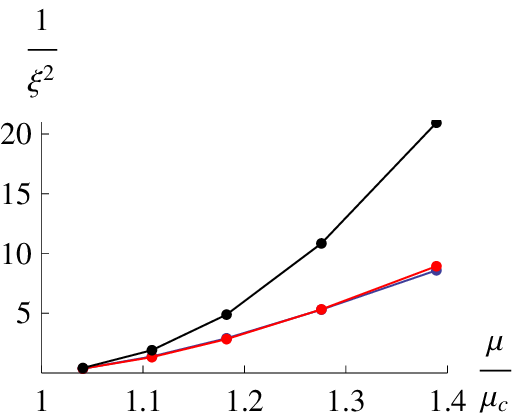}
\caption{\label{fig:scales} The length scales related to different
  operator expectation values. The different colors blue (up), black
  (middle) and red (down) correspond to the scales of the condensate,
  superfluid density and charge density, respectively. Left is for
  $\mathcal{O}_1$ and right is for $\mathcal{O}_2$. }
\end{center}
\end{figure}
We can identify a third scale by fitting the superfluid density as
\beq
\rho_s(\rho)=\rho_s(\infty)(1-\frac{\xi_s^2}{ \rho^2}+...).\label{eq:asyrhos}
\eeq
In Fig.(\ref{fig:scales}), the different scales as determined from the
asymptotic forms (\ref{eq:asycond}), (\ref{eq:asyrho}) and
(\ref{eq:asyrhos}) are graphed.  As was found for holographic dark
solitons, the length scales typically have different dependences on
the chemical potential, indicating that there are independent scales
governing the physics.  There is one exception to this result, for the
$\mathcal{O}_2$ superfluid, the condensate and charge density length
scales coincide.


Another interesting quantity is the ratio of the condensate length
scale determined from the core to the one determined from the tail
$\xi_1/\xi_2$. It was noted in \cite{Randeria} that for a BCS
superfluid, near zero temperature, these two length scales are very
different. The basic idea behind this is that the physics at the
vortex core is determined by the normal phase properties of the
superfluid, so that the length scale is related to the microscopic
Fermi momentum, $\xi_1\sim1/k_F$, while the length scale in the vortex
tail is determined by the physics of the superfluid phase, where the
relevant scale is the inverse gap $\xi_2\sim1/\Delta$. Thus, in weakly
coupled BCS theory $\xi_1\ll \xi_2$. For a more tightly bound
superfluid it was found in  \cite{Randeria} that the two length scales
coincide at the unitarity limit (infinite scattering length) and on
the BEC side of the BCS-BEC crossover.\footnote{We found in section
  \ref{sec:GP} that the ratio of the two length scales is not one, but
  this is not really to be expected, but rather both of them are
  proportional to the same length scale $\xi=1/\sqrt{\mu^2-V}$.}

\begin{figure}
\begin{center}
\includegraphics[scale=1]{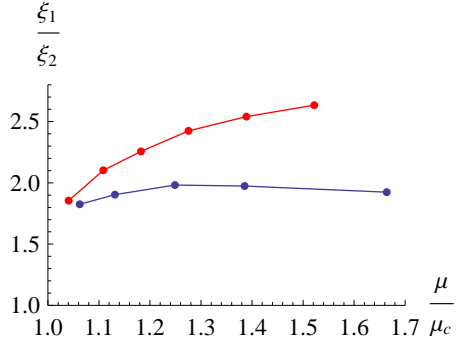}
\caption{\label{fig:xiratio} The ratio of the core length scale to the
  tail length scale of the condensate as a function of the chemical
  potential. $\mathcal{O}_1$ is blue and $\mathcal{O}_2$ is red.}
\end{center}
\end{figure}

Our results for the ratio $\xi_1/\xi_2$ in a holographic superfluid
are shown in Figure \ref{fig:xiratio}.
There are two features worth noting.  First, for the $\mathcal{O}_1$
superfluid the ratio of tail to core length scales is consistent with
a constant ratio at low temperatures (large $\mu$).   As for the GP
equation, we estimate the core length scale using the slope of the
condensate at the origin.  There is a relatively large uncertainty in
the finding the slope in the core because the spatial gradients are
large (especially at large $\mu$).  The uncertainty in the length
scales may be estimated as in \cite{Keranen:2009ss}.  We find that the
uncertainty ranges from $1-10\%$ with increasing chemical potential.
For the $\mathcal{O}_2$ superfluid the ratio of length scales clearly
depends on the chemical potential, consistent with there being two
distinct length scales in the vortex's core and tail.

The second feature we note is that the ratio of the length scales is
larger for $\mathcal{O}_2$ than for $\mathcal{O}_1$.  Given that other
quantities (including the density deplitions and the critical
velocities) indicated that $\mathcal{O}_2$ might be a BCS-like
superfluid.  This is a surprising feature.  In a speculative vein, if
we assume that the $\mathcal{O}_2$ liquid is a fermionic superfluid,
we would interpret the absence of a small core length scale as
suggesting a vanishing Fermi momentum. This is also supported by the
fact that we do not see any Friedel oscillations in our solutions,
which would have a wavelength proportional to $k_F$ \cite{antezzaetal,
  Randeria}. One possibility is that, there are a large number of
charged fermion species $N$ contributing to the condensate. In this
way, one could put an infinite charge to the ground state without
having to occupy higher energy states (assuming the charge density is
finite), which would mean that the Fermi surface, and thus the Fermi
momentum would scale to zero as $N\rightarrow\infty$.

\section{Density depletion}\label{sec:depletion}

An interesting observable in the vortex solutions is the charge
density profile. It carries non-trivial information about the fraction
of the charged matter in the condensate. Near $T=T_c$ the density
depletion in the core of the vortex is very small, which suggests that
the part of charged matter in the condensate is small compared to the
total charge density. This is found for both condensing operators
$\mathcal{O}_1$ and $\mathcal{O}_2$.

A pronounced difference between $\mathcal{O}_1$ and $\mathcal{O}_2$
becomes clear when the temperature is lowered. For $\mathcal{O}_2$ the
density depletion fraction seems to be saturating at 40$\%$ while for
$\mathcal{O}_1$ it is likely to grow near 100$\%$.  This same basic
pattern was also observed for holographic dark solitons
\cite{Keranen:2009vi,Keranen:2009ss}.  Comparing the density
depletions to those obtained in a non-relativistic setting
\cite{Randeria} suggests that one may identify $\mathcal{O}_2$ as a
BCS type superfluid and $\mathcal{O}_1$ as a BEC type superfluid.
\begin{figure}
\begin{center}
\includegraphics[scale=1]{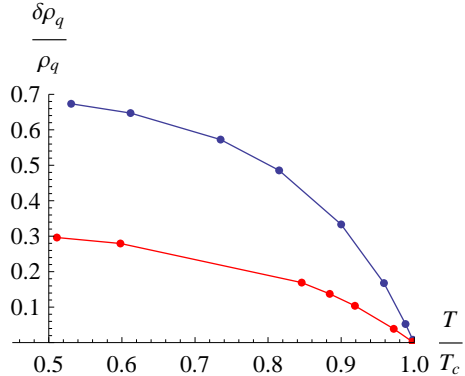}
\caption{\label{fig:dep} The density depletion fraction as a function
  of temperature. The upper curve is $\mathcal{O}_1$ (blue) while the
  lower curve is $\mathcal{O}_2$ (red).}
\end{center}
\end{figure}
\section{Critical velocity}\label{sec:velocity}
An interesting quantity in the context of superfluids is the critical
superfluid velocity, above which the superfluidity of the system is
destroyed and the flow of the fluid starts to dissipate.
\subsection{Landau criterion}

At low temperature the critical superfluid velocity can be estimated
by the Landau criterion, which goes as follows. Consider the
superfluid moving in a container with a velocity \textbf{v}.
\begin{figure}[h]
\begin{center}
\includegraphics[scale=1]{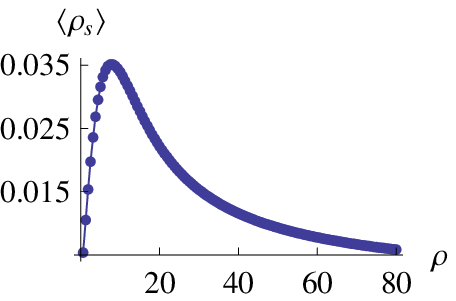}
\includegraphics[scale=1]{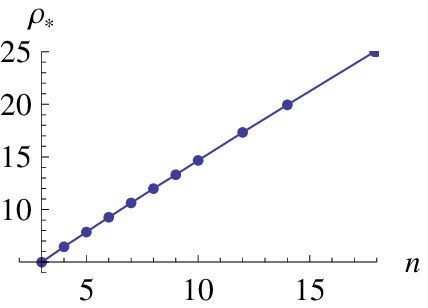}
\caption{\label{fig:critical} The left figure shows a typical profile
  for the supercurrent in a vortex. The right figure shows the
  critical radius for different values of $n$ and it seems to be a
  linear function.}
\end{center}
\end{figure}
Dissipation occurs when the wall of the container or some other defect
starts to excite quasiparticles in the fluid\footnote{For the
  dissipation one does not need a container, it is enough to have
  relative motion between superfluid and normal components.}. Let us
consider a quasiparticle with energy $\varepsilon_p$ and spatial
momentum $\textbf{p}$. Obviously these together form the momentum four
vector. As we boost back to the rest frame of the container, the time
component of the four momentum transforms into
\beq
\varepsilon'=\frac{\varepsilon_p+\textbf{p}\cdot \textbf{v}}{\sqrt{1-v^2}}.\label{eq:land1}
\eeq
Whenever the energy of the quasiparticle $\varepsilon'$ is negative,
as measured from the rest frame of the container, it becomes
energetically favorable to create such excitations. The expression
(\ref{eq:land1}) is minimized when $\textbf{p}$ and $\textbf{v}$ are
antiparallel. This gives us the Landau critical velocity
\beq
v_{crit}=\textrm{min}_p\frac{\varepsilon_p}{|\textbf{p}|},
\eeq
where the minimum is taken over all possible quasiparticle
excitations. This shows that it is not necessary for the quasiparticle
spectrum to be gapped, but there can be excitations with linear
dispersion relation for small momenta.

\subsection{Determination of the critical velocity from the vortices}
We can determine a critical velocity from the vortices as follows. We
begin by asking a simple question: Why does the radius of a vortex
increase as the winding number is increased? The simplest physical
reason is that the velocity of the superfluid flow around the vortex
increases as $v\sim n/\rho$ and at some radius $\rho_{*}$ the
superfluid velocity gets larger than the critical velocity and inside
that radius the condensate vanishes since it is no longer
energetically favorable.

Within our numerical solutions we can easily test this idea, by
plotting the critical radius $\rho_{*}(n)$ and seeing whether it
behaves as a linear function of the winding $n$, as should be if the
idea of critical velocity is the correct physical reason for the grow
of the vortices. Following ideas from \cite{Randeria}, we define the
critical radius to be the point where the current $j^{i}(\rho)$
reaches its maximum. A typical profile of the supercurrent $j_s(\rho)$
for a vortex is shown in Fig.(\ref{fig:critical}). Inside the a
critical radius, where the supercurrent reaches its maximum value, the
superfluid density starts to decrease causing the supercurrent to
decrease. The critical radius as a function of the winding number,
$\rho_{*}(n)$ seems to be indeed a linear function of $n$ as shown in
Fig.(\ref{fig:critical}).

\begin{figure}
\begin{center}
\includegraphics[scale=1]{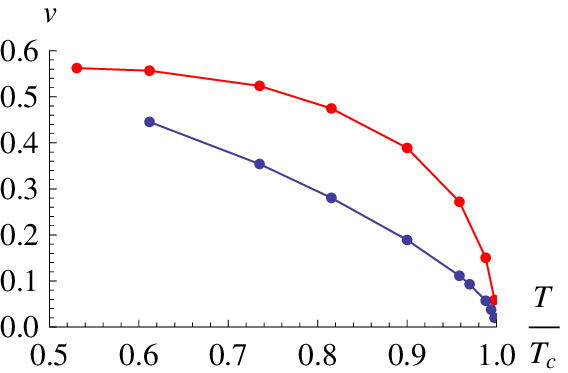}
\includegraphics[scale=1]{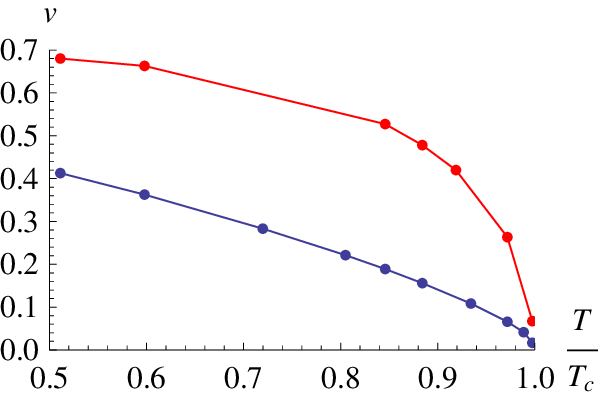}
\caption{\label{fig:crito1} The upper curve is the sound velocity and the
lower curve is the critical velocity as a function of the temperature as
determined from the vortices. $\mathcal{O}_1$ is left and $\mathcal{O}_2$ is right.}
\end{center}
\end{figure}

By generating vortex solutions with a high winding number $n=20$, for
different values of the chemical potential (or equivalently, for
different values of the temperature) we can see the behavior of the
critical velocity as a function of the chemical potential. The results
are shown in Fig.(\ref{fig:crito1}) for different condensing operators
$\mathcal{O}_1$ and $\mathcal{O}_2$.

Near the critical temperature the functional form of the critical
velocity fits well with a square root behavior
\beq
v_c\approx v_0\sqrt{1-\frac{T}{T_c}}.
\eeq  This is the same functional form as was found in \cite{Herzog:2008he}.
It is interesting to compare the critical velocity determined by the
vortices to the Landau critical velocity. For bosonic superfluids and
tightly bound fermionic superfluids, the quasiparticle determining the
Landau critical velocity is a collective mode (sound mode)
\cite{randeria2}.  A vortex's critical velocity depends on the type of
superfluid.  For a weakly coupled fermionic superfluid it is the
breaking of Cooper pairs that determines the critical velocity, while
for a bosonic superfluid a sound mode sets the relevant scale. In
\cite{HerzogYarom} it is found that the second sound is the smallest
(known) sound velocity.  Thus, it should determine the Landau critical
velocity for a bosonic superfluid.  We can use this to probe the
superfluid type.  In Fig.(\ref{fig:crito1}), the critical velocities
are plotted with the corresponding sound velocities.

At low temperatures (large $\mu$), for the  $\mathcal{O}_1$
superfluid, the critical velocity is closer to the sound velocity than
for the $\mathcal{O}_2$ superfluid.  This indicates that the vortex
critical velocity is more likely to be set by a sound mode for the
$\mathcal{O}_1$ than for $\mathcal{O}_2$.

\section{Free energy and a puzzle with the BKT argument}\label{sec:bkt}
\begin{figure}
\begin{center}
\includegraphics[scale=1.2]{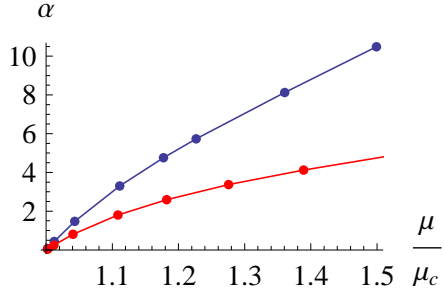}
\caption{\label{fig:alpha} Free-energy coefficients as functions of
  the chemical potential. Blue (up) is $\mathcal{O}_1$ and red (down)
  is $\mathcal{O}_2$.}
\end{center}
\end{figure}

Next, we will show that the vortex solutions have a logarithmically
divergent free energy as is usual in superfluids (for a related
calculation see \cite{Montull:2009fe}). According to the AdS/CFT
dictionary the free energy of the boundary QFT is identified with the
Euclidean on shell action in the bulk AdS. First we will evaluate the
Lorenzian on-shell action on the vortex solution, with a cutoff
$z=\epsilon$ at the AdS boundary and an infrared cutoff at the radial
position $\rho=R_{c}$.
\begin{align}
S_{AdS}&=\int
dt\int_{\epsilon}^{1}dz\int_0^{2\pi}d\theta\int_{0}^{R_c}\rho
d\rho\Big(-\frac{f}{z^2}|\partial_z\Psi|^2-\frac{1}{z^2}|\partial_{\rho}\Psi|^2\nonumber
\\
&+\frac{1}{fz^2}A_{t}^2|\Psi|^2-\frac{m^2}{z^4}|\Psi|^2-\frac{1}{2\rho^2}(\partial_\rho
A_{\theta})^2+\frac{1}{2}(\partial_z A_t)^2\nonumber
\\
&+\frac{1}{2f}(\partial_{\rho}A_t)^2-\frac{f}{2\rho^2}(\partial_z
A_{\theta})^2-\frac{1}{z^2\rho^2}
|\partial_{\theta}\Psi-iA_{\theta}\Psi|^2\Big)\label{eq:E}
\end{align}
After substracting a counterterm $S_{ct}=\int d^3x|\Psi|^2/\epsilon^3$
the Lagrangian density is finite everywhere for the vortex solutions,
but the action diverges due to integral over $\rho$. Thus, leading
terms in the free energy are the diverging ones as
$R_c\rightarrow\infty$. We will concentrate here on these terms.
Because the diverging terms are a large $\rho$ effect, they may be
captured using the asymptotic expansion in Appendix \ref{sec:asy}
\beq
\tilde{R}=\tilde{R}^{0}(z)+\frac{\delta\tilde{R}(z)}{\rho^2},\quad
A_t=A_t^{0}(z)+\frac{\delta A_t(z)}{\rho^2},\quad
A_{\theta}=A_{\theta}^{0}(z)+\frac{\delta
  A_{\theta}(z)}{\rho^2}.\label{eq:expansion}
\eeq
At leading order in this expansion the Lagrangian density is simply
$\mathcal{L}(\tilde{R}=\tilde{R}^{0}, A_t=A_t^{0},
A_{\theta}=0)$. This gives rise to an extensive term in the free
energy diverging as $R_c^2$, which is exactly the free energy of the
translationally invariant symmetry breaking state. At subleading order
there are logarithmic divergences in the free energy. These have two
sources, the two last terms in (\ref{eq:E}) are log divergent when
evaluated on $A_{\theta}^{0}(z)$, while the power law correction to
$A_{\theta}$ in (\ref{eq:expansion}) gives finite subleading
terms. Another source of possible log divergences are the first order
corrections to $\tilde{R}$ and $A_t$ in (\ref{eq:expansion}). It is
easy to see that these terms have to vanish, since they give rise to
terms that are proportional to the asymptotic equations of motion
(\ref{eq:taileq}), schematically
\beq
\int d^4x\Big(\frac{\delta S}{\delta A_t}\frac{\delta
  A_t(z)}{\rho^2}+\frac{\delta S}{\delta \tilde{R}}\frac{\delta
  \tilde{R}(z)}{\rho^2}\Big),
\eeq
and to boundary terms $\delta A(0)\partial_z A^{0}_t(0)$ and
$\delta\tilde{R}(0)\partial_z\tilde{R}(0)$, which both are
vanishing. Thus, we see that the difference between action evaluated
on the translationally invariant symmetry breaking solution and the
vortex is given by
\beq
\Delta S=-2\pi\textrm{log}\Big(\frac{R_c}{\xi}\Big)\int dt
dz\Big[\frac{1}{2}(\tilde{R}^{0})^2(n-A_{\theta}^{0})^2+\frac{f}{2}(\partial_z
  A_{\theta}^{0})^2\Big],\label{eq:FreeE2}
\eeq
where we have neglected terms that are finite in the limit
$\rho\rightarrow\infty$. By using the equations of motion of the
$A_{\theta}$ field we see that the integrand in (\ref{eq:FreeE2})
becomes simply $n\partial_z(f\partial_z A_{\theta})$. By continuing
the action to Euclidean time with period $\beta=1/T$ we end up with
the free energy difference
\beq
\Delta\Omega=\alpha\textrm{log}\Big(\frac{R_c}{\xi}\Big),\label{eq:FeeE3}
\eeq
where $\alpha=\pi n (\partial_z A_{\theta})|_{z=0}$. The coefficient
$\alpha$ is shown in figure \ref{fig:alpha} as a function of the
chemical potential.
In the section \ref{sec:scales} we showed that the superfluid density
for the vortex solution is simply $\rho_s=\mu
(\partial_zA_{\theta})/n$. Thus, we see that the vortex free energy is
given by
\beq
\Delta\Omega=\frac{\pi\rho_s(\infty) n^2}{\mu}\textrm{log}\Big(\frac{R_c}{\xi}\Big),
\eeq
where $\xi$ is a length scale which measures the vortex core size. The
vortex free energy agrees exactly with the one determined from
symmetry arguments in section \ref{sec:Goldstone}, as it
should. Furthermore, $\alpha/2\pi$ is the coupling constant in front
of the Goldstone effective action for the holographic model.

There is a slight puzzle involved with the free energy as it vanishes
at $T=T_c$. Even though we have calculated the vortex free energy, we
are missing a piece involving the vortex entropy, since we have fixed
the center of the vortex. The entropy of a single vortex is simply the
number of possible vortex states. Since the vortex occupies an area of
order $\xi^2$, the number of possible vortex positions is
$R_c^2/\xi^2$ and thus, the entropy is
$S=\textrm{log}(R_c^2/\xi^2)$.\footnote{Rather than appeal to this
  heuristic derivation of the entropy one can see the same term
  arising when one path integrates the moduli associated with
  translating the vortex core in the $\rho-\theta$ plane.} The free
energy of a single vortex is
\beq
F=E-TS=(\alpha-2T)\textrm{log}(\frac{R_c}{\xi}).
\eeq
This suggests that at some temperature $below$ the critical
temperature of the phase transition described in \cite{Gubser:2008px},
it becomes entropically favorable for a
Berezinskii-Kosterlitz-Thouless (BKT) transition.  In such transitions
it is entropically favorable for vortex/anti-vortex pairs to
deconfine, destroying the superfluid state.  A potential pitfall of
the above argument is that the semi-classical gravity approximation
involves a large-$N$ limit.  There is the possibility of an order of
limits issue when taking both $N$ and $q\ \rightarrow\infty$.  It is
possible that factors of $N$ enter into the ratio $E/T$ but do not
effect the entropy $S$.\footnote{$S$ is determined by the metric on
  moduli space and its leading contribution is independent of coupling
  constants in the gravitational theory}  In this case there would be
an order of limits issue, which might drive the BKT transition
temperature to the superfluid transition temperature.  To clarify this
issue it may be necessary to include $1/N$ corrections.  We will save
a more detailed study for later work.  It would also be interesting to
see where the potential BKT transition is placed in the phase diagrams
of \cite{Japanesegroup}.

\section{Discussion}\label{sec:discussion}
In this work we have studied vortex configurations permitted in
holographic superfluids.  These vortex solutions are characterized by
a depletion of the condensate and charge density in the vortex core
and the quantization of the circulation around the vortex.  In the
bulk these solutions correspond to vortex configurations in an
Einstein-Maxwell-Higgs system, which have vanishing boundary values
for the spatial components of the vector field.  Bulk fluxes do not
become boundary gauge fields, but instead map into the charge and
superfluid densities in the boundary theory.

We explored features of the vortices as functions of the chemical
potential and type of condensing operator.  We have used the fact that
vortex configurations are not small deviations from the homogeneous
condensate to probe both long and short distance features of
holographic superfluids.  There has been much work interpreting the
system of \cite{Hartnoll:2008vx} as a superconductor.  This may be
appropriate for features which do not require a dynamical gauge field
(such as conductivity calculations and the superconducting gap).
However, for other features, such as the Meissner effect, a dynamical
gauge field and Maxwell's equations are required.  Then, vortices act
as sources for the gauge field and require magnetic fluxes.  However,
the existence of vortex solutions even in the absence of a magnetic
field on the boundary indicates that these systems may more likely be
a superfluid or a thin film superconducting states.

The key feature of a thin superconducting film of thickness $d$ is
that the magnetic fields, as they live in $3$ spatial dimensions,
spread out in space outside the film and thus make the penetration
depth enhanced into $\lambda_{eff}=\lambda^2/d\gg\lambda$. For a thin
film $\lambda_{eff}$ can be really large (for $d\approx 10 - 100 Å$
Angstroms, $\lambda_{eff}\approx\mathcal{O}(1 cm)$), and the dynamics
of vortices in the system is equivalent to that of a superfluid.  It
becomes possible to see a BKT phase transition in a
superconductor. This enhancement was noted in \cite{Pearl:1964} and
for a nice review see \cite{Nelson:1995}. Now for a single vortex that
has a magnetic flux of a one flux quantum $\Phi_0$, the magnetic field
is spread on a large area $A\sim\lambda_{eff}^2$. Thus the magnitude
of the magnetic field inside the vortex core is of the order
$B\sim\Phi_0/A$,\footnote{A similar scaling of the magnetic field with
  the area was used in \cite{Montull:2009fe}, but there the area $A$
  is kept finite and thus, the external magnetic field is nonzero.}
which is practically zero. Thus, when modeling vortices in a thin
superconducting film it is a good approximation to set the local
magnetic field to zero.

The specific system studied here is a relativistic superfluid, but the
features of non-relativistic superfluid vortices are a useful guide
for interpreting our results.  Specifically, we find that vortex
configurations have distinct length scales characterizing the
variations of the condensate, charge density, and superfluid density.
Each of these length scales is characterized by a
$1/\sqrt{\mu/\mu_c-1}$ near the critical point.  However, each of
these scales generically has a different dependence on the chemical
potential for larger $\mu$ as is seen in Fig.(\ref{fig:scales}).

As with holographic dark solitons \cite{Keranen:2009vi,Keranen:2009ss}
the behavior of the charge density depletion fraction of the vortex
core is very different for the $<\mathcal{O}_1>$ and $<\mathcal{O}_2>$
condensates.  For both dark solitons and vortices, the
$<\mathcal{O}_1>$ condensate has near $100\%$ charge depletion in the
core for low temperature (large $\mu$).  For $<\mathcal{O}_2>$ the
core density depletion is much more modest, near $40\%$.  This is
consistent with the picture that the $<\mathcal{O}_1>$ condensate is
BEC-like, comprised of a point-like boson.  Similarly, this is
suggestive that the $<\mathcal{O}_2>$ condensate is more BCS-like,
comprised of non-local Cooper pairs.

In \cite{Randeria} the impact that varying the type of superfluid
across the BEC-BCS crossover has on vortices was explored.  In that
non-relativistic system, it was found that for BEC superfluids there
is a single length scale characterizing the variations of the
condensate.  However, for a BCS superfluid the condensate profile was
characterized by two length scales.  In the vortex's tail the
characteristic length scale is set by the size of the gap.  Near the
core, as the system locally approaches the normal phase, $k_F$
determines the characteristic scale.  In Fig (\ref{fig:xiratio}) we
see some evidence that similar things happen for holographic
superfluids.  Specifically we see that there is really one length
scale characterizing both the core and tail region for an
$<\mathcal{O}_1>$ type condensate.  For $<\mathcal{O}_2>$ the core and
tail are characterized by two distinct scales.  However, the picture
is not quite as clean as one might have hoped.  In the
non-relativistic system, the core length scale is smaller than for the
tail.  In the holographic $<\mathcal{O}_2>$ superfluid we find the
opposite.  If one accepts the basic picture that that the two types of
holographic superfluids are of BEC and BCS types, on might view the
core length scale as giving information about the Fermi surface (in
the large $N$ limit).  Specifically, we would conclude that the Fermi
momenta is vanishing in the large $N$ limit.  The vanishing of $k_f$
is consistent with the absence of Friedel oscillations in vortices and
dark solitons \cite{Keranen:2009vi,Keranen:2009ss}.  For other recent
discussions of Fermi surfaces in holographic quantum liquids see
\cite{Chen:2009pt,Faulkner:2009am,Gubser:2009dt}.  It would be
interesting to compare their results to the behavior we see in the
vortex cores.

We also study the manner in which vortices allow one to estimate the
critical velocity of the superfluid. Heuristically, this is due to the
fact that the superfluid's local velocity increases beyond its
critical velocity as one approaches the core.  We can then estimate
the critical velocity by the radius where the superfluid density
starts to drop off.  At low temperatures the Landau criterion sets an
upper bound for the critical velocity of the superfluid in terms of
the lowest sound mode.  The fact that the $\mathcal{O}_1$ superfluid's
critical velocity comes closest to saturating the Landau criterion
indicates that it is more likely to be a BEC-like superfluid (whose
critical velocity is set by a sound mode).

In addition to studying the vortices scales of variation, we confirmed
that their energy cost over the homogeneous solution diverges
logrithmically.  While this is generically for any $2+1$ dimensional
vortex, the holographic reason is that the exterior of an
AdS-Schwarzschild black hole effectively acts as a finite box in the
radial direction.  This is a nice confirmation that the bulk vortices
give rise to superfluid vortices in the boundary theory as opposed to
superconductor vortices.  On general grounds one expects vortices in a
superconductor to be finite energy excitations.

In addition it appears that one can have a BKT transition at
temperatures below the temperature of spontaneous symmetry breaking.
To definitively resolve this issue on may need to include $1/N$
effects as well as gravitational backreaction.  As in
\cite{Keranen:2009vi,Keranen:2009ss}, the absence of gravitational
back reaction limits our ability to take the zero temperature limit of
vortex configurations.  Including gravitational backreaction would
allow us to see many features unobstructed by thermal fluctuations.
In addition, the backreacted geometry would present a very novel
example of black hole hair.  For recent work on black holes with
inhomogeneous hair see \cite{Nakamura:2009tf}.

Throughout this paper and in studies of dark solitons we have
repeatedly seen features suggesting that the the choice of
quantization in $AdS_4$ corresponds to two types of superfluids.  It
would be worthwhile to probe this feature more closely.  Ideally one
would like to understand if there is any holographic analog of the
"unitarity" limit in the BEC-BCS crossover.  One way to probe this
physics might be to vary the $m^2$ parameter in the bulk Lagrangian (studies in this direction were performed in \cite{Horowitz:2008bn,Umeh:2009ea,Kim:2009kb}).
If we associate the operator scaling dimension with the type of
superfluidity, it is interesting to speculate that the unitarity limit
would correspond to saturating the BF bound, $m^2=-9/4$, in $AdS_4$.
One way that one might try to see this would be to study the way that
the critical velocity varies with $m^2$.  In \cite{Randeria} it was
shown that the critical velocity was not a monotonic function across
the BEC-BCS crossover. The peak value occurrs in the unitarity
limit. Since the system is relativistic one might be able to go all
the way from BCS to BEC and to relativistic Bose-Einstein condensation
(RBEC) \cite{Nishida:2005ds}.

Finally, it would be very interesting to explore all of these
questions in a setting more appropriate for laboratory experiments.
Specifically we anticipate all of these ideas have analogs in
non-relativistic gauge-gravity duals.  It would be very interesting to
study vortices in the backgrounds described in
\cite{Son:2008ye,Balasubramanian:2008dm,Kachru:2008yh,Brynjolfsson:2009ct}

\bigskip

{\bf Acknowledgements} We would like to thank Clifford Johnson, Rob
Leigh, Makoto Natsuume, and Kari Rummukainen for useful
comments. V.K. and E.K-V. have been supported in part by the Academy
of Finland grant number 1127482.  E.K-V. and S.N. thank the Aspen
Center for Physics for their hospitality during the early stages of
this work.

\appendix
\section{Gauge fixing}\label{sec:gauge}

In this appendix we will outline gauge fixing necessary to obtain the
equations in section \ref{sec:model}.  Because vortex solutions are
cylindrically symmetric in the $(\rho,\theta)$ plane, we can take all
gauge invariant quantities independent of the angle $\theta$. Since we
are interested in static solutions, we will also take all the fields
to be independent of the time coordinate.\footnote{More general
  condition would have been to take all the gauge invariant quantities
  independent of $t$, but this leads to trouble discussed in
  \cite{Gubser:2008px}} Choosing the gauge, $A_z=0$, and using
$\partial_{\theta}R=0$ (since $R$ is gauge invariant) equations
(\ref{eq:ELeq1}), (\ref{eq:ELeq2}) and (\ref{eq:ELeq3}) become
\begin{align}
0 &= z\partial_z(\frac{f}{z^2}\partial_z
R)+\frac{m^2}{z^2}(\frac{R}{z})+\frac{1}{\rho}\partial_{\rho}(\rho\partial_{\rho}
(\frac{R}{z}))\nonumber
\\
&-(\frac{R}{z})(-\frac{1}{f}A_t^2+f(\partial_z\chi)^2+\frac{(A_{\theta}
-\partial_{\theta}\chi)^2}{\rho^2}+(A_{\rho}-\partial_{\rho}\chi)^2)
\\
0 &= \frac{1}{\rho}\partial_{\rho}(\rho\partial_z
A_{\rho})+\frac{1}{\rho^2}\partial_{\theta}\partial_z
A_{\theta}-(\frac{R}{z})^2\partial_z\chi\label{eq:zeq}
\\
0 &= f\partial^2_z A_t+\frac{1}{\rho}\partial_{\rho}
(\rho\partial_{\rho} A_t)+\frac{1}{\rho^2}\partial_{\theta}^2 A_t-(\frac{R}{z})^2 A_t
\\
0&=\partial_z(f\partial_z
A_{\rho})+\frac{f}{\rho^2}\partial_{\theta}^2A_{\rho}
-\frac{f}{\rho^2}\partial_{\theta}\partial_z
A_{\theta}-(\frac{R}{z})^2
(A_{\rho}-\partial_{\rho}\chi)\label{eq:rhoeq}
\\
0 &= \partial_z( f\partial_z
A_{\theta})+\rho\partial_{\rho}(\frac{1}{\rho}\partial_{\rho}
A_{\theta})-\rho\partial_{\rho}(\frac{1}{\rho}\partial_{\theta}
A_{\rho})-(\frac{R}{z})^2
(A_{\theta}-\partial_{\theta}\chi)
\\
0 &= \partial_z(f(\frac{R}{z})^2\partial_z\chi) -
\frac{1}{\rho}\partial_{\rho}(\rho(\frac{R}{z})^2(A_{\rho}-\partial_{\rho}\chi))\nonumber
\\
&-\frac{1}{\rho^2}\partial_{\theta}((\frac{R}{z})(A_{\theta}
-\partial_{\theta}\chi)).\label{eq:chieq}
\end{align}
Furthermore, to model a vortex we want the radial current to vanish,
that is, $A_{\rho}-\partial_{\rho}\chi=0$.

We may use the residual gauge invariance and the requirements of
regularity along the horizon and in the vortex core to simplify the
remaining equations.  Using
$\partial_{\theta}(\partial_{\theta}\chi-A_{\theta})=0$, equation
(\ref{eq:chieq}) implies
\beq
\partial_{z}((\frac{R}{z})^2f(z)\partial_{z}\chi)=0.
\eeq
Which has the solution
\beq
\partial_z\chi=\frac{z^2C(\rho)}{R^2f(z)}.
\eeq
We can note that $C$ is independent of $\theta$ since
$0=\partial_{\theta}(\partial_z\chi-A_z)=\partial_{\theta}\partial_z\chi$.
Regularity of $\chi$ at the horizon forces us to set $C(\rho)=0$.

Cylindrical symmetry implies that $\partial_{\theta}F_{\rho\theta}=0$
and $\partial_{\theta}F_{\theta z}=0$. Substituting these in
(\ref{eq:rhoeq}), gives
\beq
\partial_z(f\partial_zA_{\rho})=0,
\eeq
which has a solutions $\partial_z
A_{\rho}=\frac{D(\rho,\theta)}{f(z)}$. Again, regularity at the
horizon requires $D=0$. This means that
$A_{\rho}=A_{\rho}(\theta,\rho)$. Thus, by a gauge transformation we
can set $A_{\rho}=0=\partial_{\rho}\chi$

The symmetry also imposes $\partial_{\theta}F_{z
  \theta}=\partial_{\theta}\partial_z A_{\theta}=0$, This has the
general solution
$A_{\theta}=A_{\theta}^{(1)}(\theta,\rho)+A_{\theta}^{(2)}(z,\rho)$. From
$\partial_{\theta}(\partial_{\theta}\chi+A_{\theta})=0$ and
$\partial_{\rho}\chi=0$ we can conclude that $A_{\theta}^{(1)}$ is
independent of $\rho$ and can be thus absorbed into $\chi$ with a
gauge transformation. Thus $A_{\theta}=A_{\theta}(z,\rho)$.

Similarly, cylindrical symmetry also requires
$0=\partial_{\theta}(\partial_{\theta}\chi+A_{\theta})=\partial_{\theta}^2\chi$,
which implies that $\chi=\alpha+\beta \theta$ for $\alpha$ and $\beta$
constants. In order for $\Psi$ to be a single valued function $\beta$
has to be an integer $n\in\mathbb{Z}$, which is the winding number of
the vortex. We will also set $\alpha=0$ as a part of the gauge choice.
Note that the field $\chi(z,x)$ is completely determined by the
equations of motion, and thus, cannot be thought of as a
non-normalizable mode.\footnote{In Appendix \ref{sec:appendix1} we
  show that for a general solution of the field equations
  (\ref{eq:ELeq1}), (\ref{eq:ELeq2}) and (\ref{eq:ELeq3}), the phase
  field $\chi$ is completely determined by the equations of motion and
  there are no freedom to choose boundary conditions for $\chi$ at the
  AdS boundary.}  The final simplification is due to the observation
that $A_{t}-\partial_0\chi=A_{t}$ is gauge invariant, and therefore is
independent of $\theta$

Taking advantage of the cylindrical symmetry, the equations may be
written
\begin{align}
0&= z\partial_z(\frac{f}{z^2}\partial_z
R)+\frac{m^2}{z^2}(\frac{R}{z})+\frac{1}{\rho}\partial_\rho(\rho\partial_\rho
(\frac{R}{z}))\nonumber
\\
&-(\frac{R}{z})(-\frac{1}{f}A_t^2+\frac{(A_\theta-n)^2}{\rho^2})
\\
0 &= f\partial^2_z A_t+\frac{1}{\rho}\partial_\rho(\rho\partial_\rho A_t)-(\frac{R}{z})^2A_t
\\
0&= \partial_z( f\partial_z
A_\theta)+\rho\partial_\rho(\frac{1}{\rho}\partial_\rho A_\theta)
- (\frac{R}{z})^2(A_\theta-n).
\end{align}
After defining $\tilde{R}=R/z$, these equations are equivalent to
(\ref{eq:eqr}-\ref{eq:eqs}).

\section{Boundary conditions for the $\chi$ field}\label{sec:appendix1}
According to the AdS/CFT dictionary the expectation value of the order
parameter field is given by
\beq
\langle \mathcal{O}_i\rangle\sim e^{i\chi(0,x)}R_{(i)}(0,x).
\eeq
Thus, $\chi(0,x)$ should be a quantity determined by the state one is
looking at. Therefore one should not be able to choose the value of
$\chi(0,x)$ as a boundary condition. We will show in this appendix
that this is indeed the case for a general time independent solution
of the equations of motion. The equations of motion for $\chi$ in
terms of the $\tilde{R}$ field are in the gauge $A_z=0$ given by
\begin{align}
& \partial_z(f\tilde{R}^2\partial_z\chi) + \partial_{i}(\tilde{R}^2(\partial_i\chi -A_i))=0.\label{eq:chieq2}
\end{align}
Since (\ref{eq:chieq2}) is elliptic, boundary conditions should be
provided at every spatial boundary. Since at the horizon
(\ref{eq:chieq2}) becomes parabolic, it provides an effective boundary
condition at the horizon (this is usually referred as a regularity
condition). Next we will show that the other boundary condition at the
AdS boundary is fixed by the equations of motion for the
$\mathcal{O}_2$ condensate and by a "regularity" condition for the
$\mathcal{O}_1$ condensate.

We know the asymptotic expansion for the fields, $\tilde{R}=R_{(1)}+zR_{(2)}+...$ and $A_{\mu}=A_{\mu}^{(0)}+zA_{\mu}^{(1)}+...$. By regularity the expansion for $\chi$ should come in positive integer powers $\chi=\chi_{(0)}+z\chi_{(1)}+...$. Plugging these expansions into (\ref{eq:chieq2}) gives the following equation up to second order in the powers of $z$
\begin{align}
&2R_{(1)}R_{(2)}\chi_{(1)}+2R_{(1)}^2\chi_{(2)}+\partial_i(R_{(1)}^2(\partial_i\chi_{(0)}-A_i^{(0)}))=0.\label{eq:asy1}
\\
&6\chi_{(3)}R_{(1)}^2+2(R_{(2)}^2+2R_{(1)}R_{(3)})\chi_{(1)}+8R_{(1)}R_{(2)}\chi_{(2)}\nonumber
\\
&+2\partial_i(R_{(1)}R_{(2)}(\partial_i\chi_{(0)}-A_i^{(0)}))+\partial_i(R_{(1)}^2(\partial_i\chi_{(1)}-A_i^{(1)}))=0.\label{eq:asy2}
\end{align}
For the $\mathcal{O}_2$ condensate we have $R_{(1)}(x)=0$ and thus
(\ref{eq:asy1}) is trivially solved, while (\ref{eq:asy2}) simplifies
into $\chi_{(1)}=0$. This means that for $\mathcal{O}_2$, the equation
of motion forces the Neumann boundary condition
\beq
\partial_z\chi(z=0)=0,
\eeq
and thus there is no more freedom to give boundary conditions to the
$\chi$ field at the AdS boundary.

For the case of the $\mathcal{O}_1$ condensate we are not able to
derive such a boundary condition from the equations of motion. Rather,
we will use a physical argument, which sounds quite sensible. We
require that the flux of the electric $U(1)$ current into the $z$
direction vanishes as we approach the AdS boundary, since literally
the space ends there. This can be thought of as requiring the total
charge in the system to be conserved.\footnote{A similar argument was
  used in \cite{Aharony:1999ti} for the energy momentum tensor and
  conservation of the total energy, in order to derive the BF bound.}
So we require
\beq
\lim_{z_0\rightarrow0}\int_{z=z_0} d^2x\sqrt{g^{(2)}}J^z=0.\label{eq:current}
\eeq
The bulk current is given by the expression $J^{z}=\frac{-i}{2}
\sqrt{-g}g^{zz}(\Psi^{*}\partial_z\Psi-\Psi\partial_z\Psi^{*}-2iA_z|\Psi|^2)
= f\tilde{R}^2\partial_z\chi$. Now (\ref{eq:current}) implies that
\beq
\lim_{z_0\rightarrow0}\int_{z=z_0} d^2x\frac{f(z)}{z^2}\tilde{R}^2\partial_z\chi=0.
\eeq
This can be true only if $\partial_z\chi(z=0)=0$.

\section{Large $\rho$ behavior}\label{sec:asy}

In this appendix we outline an alternate approach to determine the
large radius behavior of vortex solutions.  The main observation is
that for large $\rho$ the partial differential equations may be
reduced to a set of coupled ordinary differential equations, which may
be solve using Mathematica's NDSolve routine.

In more detail, the gauge fixed equations of motion are:
\beqn &&\partial_z(f\partial_z \tilde{R})+\frac{1}{\rho}\partial_\rho(\rho\partial_\rho \tilde{R})-\left(z+\frac{(A_\theta-n)^2}{\rho^2}-\frac{A_t^2}{f}\right)\tilde{R}=0\\
&&f\partial^2_z A_t+\frac{1}{\rho}\partial_\rho(\rho\partial_\rho A_t)-\tilde{R}^2 A_t=0\\
&&\partial_z(f\partial_z A_\theta)+\rho\partial_\rho(\frac{1}{\rho}\partial_\rho A_\theta)-\tilde{R}^2 (A_\theta-n)=0.\eeqn
If we assume that that all of the fields become spatially homogeneous for large $\rho$, we see that the $\tilde{R}$ and $A_t$ reduce to the equations solved by \cite{Hartnoll:2008vx}.  We can then develop an expansion around this solution, \beq \tilde{R}=\tilde{R}^0+\frac{\dr (z)}{\rho^2}+...,\ \ \ A_t = A_{t}^0+\frac{\da (z)}{\rho^2}+...,\ \ \mathrm{and} \ \ A_\theta=A_{\theta}^0 + \frac{\db}{\rho^2}+...\eeq  One would like to solve for $A_{\theta}^0, \dr,\da,$ and $\db$ in terms of the homogeneous solutions.

The equations become \beqn
&& \partial_z(f\partial_zA_{\theta}^0)-(\tilde{R}^0)^2 (A_{\theta}^0-n) \nonumber\\
&&+ \frac{1}{\rho^{2}}\left(\partial_z(f\partial_z\db)-(\tilde{R}^0)^2\db -2\tilde{R}^0A_{\theta}^0\dr\right)+...=0\\
&&\left(\partial_z(f\partial_z \dr)-(z-\frac{(A_{t}^0)^2}{f})\dr\right)\nonumber\\
&&\ \ \ \ \ \ \ \ \ \ \ \ \ \ \ +\frac{2\tilde{R}^0A_{t}^0}{ f}\da-(A_{\theta}^0-n)^2\tilde{R}^0+...=0\\
&&\left(f\partial_z^2\da-\tilde{R}^2_0\da\right)-2\tilde{R}^0A_{t}^0\dr+...=0. \eeqn  Equating the coefficients of powers of $\rho$ gives  \beqn\label{eq:taileq}
 &&\partial_z(f\partial_zA_{\theta}^0)-(\tilde{R}^0)^2 (A_{\theta}^0-n) =0\\
&&\partial_z(f\partial_z\db)-(\tilde{R}^0)^2\db  -2\tilde{R}^0A_{\theta}^0\dr=0\\
 &&\partial_z(f\partial_z \dr)-(z-\frac{(A_{t}^0)^2}{f})\dr +\frac{2\tilde{R}^0A_{t}^0}{f}\da-(A_{\theta}^0-n)^2\tilde{R}^0=0\\
&&f\partial_z^2\da-\tilde{R}^2_0\da-2\tilde{R}^0A_{t}^0\dr=0 .\eeqn

In addition to these differential equations one must also impose
boundary and regularity conditions.  For the scalar field we should
impose the boundary conditions appropriate for the type of condensate,
$\partial_z \dr(0) = 0$ ($<\mathcal{O}_1>$ case) or $\dr(0)=0$
($<\mathcal{O}_2>$ case).  For $\da$ we require that the chemical
potential not be spatially varying, $\da(0)=0$.  Regularity at the
horizon requires $\da(1)=0$.  To study vacuum properties we should set
all external sources for the superflow to zero,
$A_{\theta}^0(0)=\db(0)=0$.

The most important thing to note is that the partial differential
equations reduce to a set of coupled ODE's, which may be solved using
any standard numerical differential equations solver.  We have checked
that the solutions to (\ref{eq:taileq}) determined using Mathematica's
NDSolve agree with the solutions described in Section \ref{sec:GS}.
The difference between solutions found with the two methods is at the
.1\% level (the level to which the equations are solved by
Gauss-Seidel method).  Because the asymptotic analysis is wholly
independent of the Gauss-Seidel approach, the agreement between the
two provides an independent check of the results in Section
\ref{sec:GS}.

\end{document}